\documentclass[sigconf,nonacm,balance=false,pbalance=true]{acmart}
\usepackage{multirow}

\setcopyright{none}
\graphicspath{{./}}
\hypersetup{
  pdfauthor={Ke Changke, Jian Wang, Xuan Xu, Yijing Zong, Jiqiang Liu}
}

\begin{document}
\title{NPUFHE: An Efficient CKKS Execution System for Ascend NPUs}

\author{Ke Changke}
\email{changke@bjtu.edu.cn}
\author{Jian Wang}
\author{Xuan Xu}
\author{Yijing Zong}
\author{Jiqiang Liu}
\authornote{Corresponding author.}
\email{jqliu@bjtu.edu.cn}
\affiliation{
  \institution{Beijing Jiaotong University}
  \city{Beijing}
  \country{China}
}
\renewcommand{\shortauthors}{Ke Changke et al.}

\begin{abstract}
Fully homomorphic encryption (FHE) enables computation directly on encrypted data, but its substantial computational cost hinders adoption in privacy-preserving artificial intelligence (AI) applications.
Neural processing units (NPUs) already deployed in AI infrastructure provide a practical platform for FHE acceleration.
However, after mapping the number theoretic transform (NTT) to matrix multiplications on Ascend NPUs, modular arithmetic and representation conversion on the Vector backend still incur substantial overhead at stage boundaries.
Evaluation-key material of key switching also increases storage and data-transfer costs, and whole-limb task assignment causes tail load imbalance.

We present \textsc{NPUFHE}, an end-to-end CKKS execution system optimized for Ascend NPUs. To reduce modular recovery and twiddle costs in NTT, we propose Boundary-Optimized NTT, which eliminates intra-axis modular recovery, compiles intra-axis twiddle factors into matrix constants, and introduces two-level folded modular recovery at axis endpoints.
To alleviate evaluation-key storage and transfer pressure, we design Compact Shoup encoding, which halves the key representation through local coefficient recovery.
Finally, we propose Cost-Aware Parallel Tiling to mitigate tail load imbalance caused by whole-limb task assignment.
It uses a cost model to adaptively partition tail tasks and smooth stepwise latency increases.

We integrate the above designs into OpenFHE.
On one Ascend 910B4, full-packing CKKS bootstrapping with real evaluation keys takes 0.95\,s, encrypted CIFAR-10 ResNet-20 inference with bootstrapping requires 9.87\,s per image, and encrypted MNIST inference achieves an amortized latency of 19.81\,ms per image.
The peak NTT throughput of NPUFHE reaches up to $184.22\times$, $2.04\times$, and $4.46\times$ that of OpenFHE on CPU, TensorFHE, and CROSS, respectively.

\end{abstract}
\keywords{Fully Homomorphic Encryption, CKKS, Number Theoretic Transform, Neural Processing Units}
\maketitle

\section{Introduction}

Fully homomorphic encryption (FHE) enables artificial intelligence (AI) services to perform computations over encrypted data without accessing the underlying plaintext.
A growing body of research applies FHE to protect sensitive data in AI applications.
CryptoFace extracts and matches facial features without exposing them~\cite{ao_CryptoFace_2025}, Blind-Touch introduces an FHE-based system for privacy-preserving fingerprint-authentication~\cite{choi_BlindTouch_2024}, and PriTran provides a CKKS-based inference system for BERT-like Transformer language models~\cite{mu_PriTran_2026}.

However, FHE computation incurs substantial latency.
CryptoFace reports an online latency of 1,364--1,446 seconds per image pair on an AWS r5.24xlarge instance~\cite{ao_CryptoFace_2025}.
An ablation study in Blind-Touch shows that evaluating its first fully connected layer on encrypted data increases total evaluation time by $23{,}224\times$, to 15,096 seconds~\cite{choi_BlindTouch_2024}.
Reducing execution overhead is therefore essential for making FHE practical.

Prior work accelerates FHE using FPGAs~\cite{agrawal_FAB_2023,agrawal_HEAP_2024,xu_FAST_FPGA_2025} and ASICs~\cite{feldmann_F1_2021,samardzic_CraterLake_2022}, as well as matrix hardware on GPUs~\cite{fan_TensorFHE_2023,fan_WarpDrive_2025,sun_TensorFHEPlus_2026} and TPUs~\cite{tong_CROSS_2026}.
We further explore Ascend neural processing units (NPUs), already deployed in AI infrastructure~\cite{zuo_CloudMatrix384_2025,epoch_AIChipSales}, to enable existing computing resources to support both plaintext AI and encrypted workloads without deploying additional FHE-specific hardware.

High matrix throughput for plaintext AI does not automatically translate into efficient ciphertext computation.
FHE execution on Ascend faces three challenges.
\noindent\textbf{(1) Vector-side boundary overhead in matrix-based NTT.}
Deeper matrix decomposition of the number theoretic transform (NTT) reduces computation but introduces additional modular recovery and representation conversion, eroding the gains of matrix acceleration.
\noindent\textbf{(2) Tail inefficiency in whole-limb task assignment.}
Underfilled final execution waves reduce resource utilization, causing stepwise latency at task-wave boundaries.
\noindent\textbf{(3) Compute--memory trade-off in modular multiplication.}
Shoup multiplication has low arithmetic cost, but its paired storage of coefficients and precomputed quotients increases the storage and transfer pressure of evaluation-key material.

To address these challenges, we present \textsc{NPUFHE}, an end-to-end CKKS execution system for Ascend NPUs.
To our knowledge, this is the first CKKS system on Ascend NPUs to support homomorphic operations, bootstrapping with real evaluation keys, and encrypted inference.
Our main contributions are as follows:

\textbf{(1) Boundary-Optimized NTT (BO-NTT).}
BO-NTT uses a bounded intermediate representation and constant compilation to eliminate intra-axis modular recovery and standalone intra-axis twiddle multiplication.
Two-level folded modular recovery restores canonical residues at axis endpoints.
Asymmetric decomposition reduces matrix work and avoids the extra data reordering introduced by decomposing both axes.

\textbf{(2) Cost-Aware Parallel Tiling (CAPT).}
CAPT retains whole-limb task assignment and tiles only the underfilled tail.
A cost model selects granularity and resource allocation to mitigate stepwise latency at task-wave boundaries.

\textbf{(3) Compact Shoup encoding.}
We store only precomputed quotients and recover coefficients when they are consumed, reducing the representation of supported evaluation-key coefficients by $50\%$.

\textbf{(4) End-to-end CKKS execution system.}
We integrate the runtime and device operators with OpenFHE.
We evaluate the system using bootstrapping with real evaluation keys, precision checks on decrypted outputs, and encrypted inference on MNIST and ResNet-20.
\section{Background}
\label{sec:background}
We review CKKS computation, Ascend NPU architecture, matrix-based NTTs, and Shoup modular multiplication.

\subsection{CKKS Computation and RNS Parallelism}
For a power-of-two polynomial degree $N$, CKKS encodes up to $N/2$ values into a polynomial, enabling approximate arithmetic on packed data~\cite{cheon_Homomorphic_2017}.
A ciphertext normally consists of two polynomials, $\mathrm{ct}=(c_0,c_1)$, with $c_0,c_1\in\mathcal R_Q=\mathbb Z_Q[X]/(X^N+1)$, where $Q=\prod_{i=0}^{L-1}q_i$ and $L$ is the number of active Q-basis residue-number-system (RNS) limbs.
Rather than operating directly modulo the large modulus $Q$, RNS-CKKS represents each ciphertext polynomial with $L$ limbs~\cite{cheon_rns_2019}.
Limb $i$ contains $N$ coefficients modulo $q_i$.
Thus, modular additions and the modular arithmetic in polynomial products can be performed in parallel across limbs.

Let $\mathrm{ct}'=(c'_0,c'_1)$ and $\mathrm{ct}''=(c''_0,c''_1)$ be ciphertexts under the same modulus $Q$ with aligned scales.
Their homomorphic sum is $\operatorname{HAdd}(\mathrm{ct}',\mathrm{ct}'')=(c'_0+c''_0,\,c'_1+c''_1)$ in $\mathcal R_Q$.
Ciphertext multiplication produces three components: $(c'_0c''_0,\;c'_0c''_1+c'_1c''_0,\;c'_1c''_1)$.
Relinearization uses key switching to convert this
three-component result into a two-component ciphertext.
Ciphertext slot rotation likewise requires key switching.
Both operations rely on precomputed evaluation-key material. This material consists of multiple RNS polynomials, and its size is jointly determined by the polynomial degree, the active set of moduli, and the key-decomposition parameters.

CKKS encodes plaintexts with scale $\Delta$, and ciphertext multiplication multiplies the input scales.
Rescaling divides by the product of the last $d$ primes and rounds, updating the modulus, scale, and active limb count:
\begin{equation}
  (Q,\Delta,L)
  \longrightarrow
  \left(
    \frac{Q}{\prod_{i=L-d}^{L-1}q_i},
    \;\frac{\Delta}{\prod_{i=L-d}^{L-1}q_i},
    \;L-d
  \right).
\end{equation}
NPUFHE uses double-prime rescaling with $d=2$, where each rescaling operation removes two physical limbs and thus consumes one logical level~\cite{kim_Reduced_2022}.
Bootstrapping restores the usable modulus budget through a sequence of homomorphic operations~\cite{kim_ARK_2022}.

\subsection{Ascend Architecture and Execution Model}
\label{sec:ascend_execution}

\begin{figure}[b]
  \centering
  \includegraphics[width=\columnwidth]{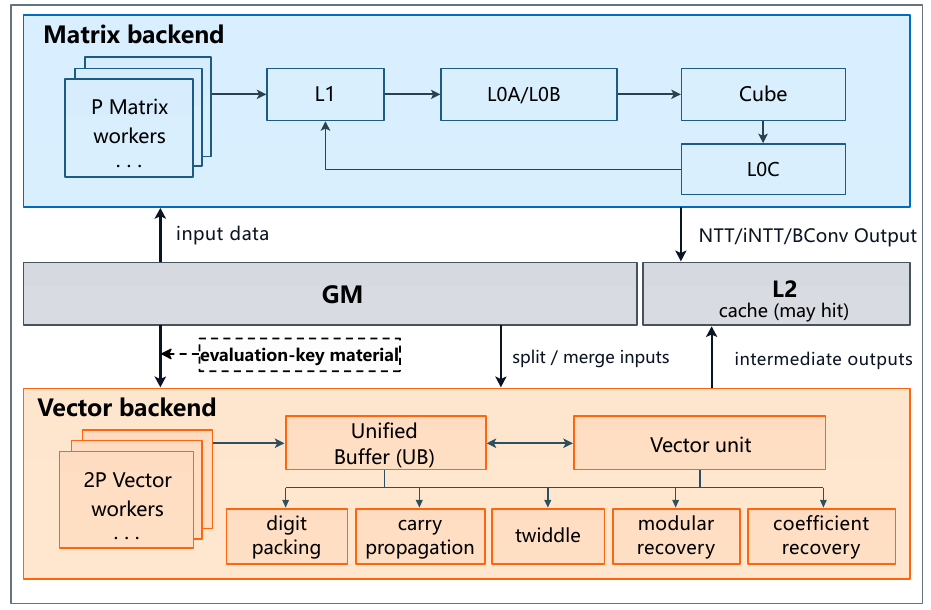}
  \caption{%
    Ascend Matrix--Vector execution model.
  }
  \Description{Ascend Matrix--Vector execution model. The two backends use independent local storage and exchange intermediate data through the GM/L2 path.}
  \label{fig:ascend_backends}
\end{figure}

As Figure~\ref{fig:ascend_backends} shows, Ascend's Matrix backend uses the Cube unit for low-precision matrix multiplication, while the Vector backend performs digit extraction, merging, and modular arithmetic~\cite{huawei_AscendC_2025}.
In Ascend NPUs, intermediate results exchanged between backends are written to global memory (GM) and subsequently reloaded.
Although these GM accesses may hit in the L2 cache, representation conversion and inter-backend data movement still incur non-negligible execution overhead.
\subsection{Matrix NTT}
\label{sec:ntt_in_ckks}

Within each RNS limb, the NTT turns negacyclic polynomial multiplication into pointwise modular multiplication.
For a prime $q$ satisfying $2N\mid(q-1)$, denote by $\psi$ a primitive $2N$-th root of unity modulo $q$ and define $\omega=\psi^2$.
The forward transform is $\hat{x}_j=\sum_{i=0}^{N-1}x_i\psi^i\omega^{ij}\pmod q$.
The inverse NTT (iNTT) uses the corresponding inverse roots of unity and the scaling factor $N^{-1}$.

Matrix-based NTTs factor the transform into smaller dense matrix products~\cite{fan_TensorFHE_2023,tong_CROSS_2026_v4}.
Let $N=N_1N_2$ and arrange the input as $X_{r,c}=x_{rN_2+c}$, where $0\le r,a<N_1$ and $0\le c,b<N_2$.
Define $F_{a,r}=(\psi^{N_2})^r(\omega^{N_2})^{ar}$, $W_{a,c}=\psi^c\omega^{ac}$, and $H_{c,b}=(\omega^{N_1})^{bc}$.
The forward NTT is $Y=\bigl((FX)\odot W\bigr)H\pmod q$, with $\hat{x}_{a+N_1b}=Y_{a,b}$ and $\odot$ denoting elementwise multiplication.
This form divides the NTT into a first-axis matrix transform, inter-axis twiddle multiplication, and a second-axis matrix transform.
Further decomposing either axis reduces matrix size but introduces additional matrix stages, twiddle multiplications, and possible intermediate data reordering~\cite{fan_TensorFHE_2023,fan_WarpDrive_2025,tong_CROSS_2026_v4}.

Low-precision matrix units cannot directly process full-width integers modulo $q$, so inputs are decomposed into $b_p$-bit digits and represented in radix $\rho=2^{b_p}$ as $x=\sum_e x_e\rho^e$, with $0\le x_e<\rho$.
Matrix multiplication accumulates digit products into planes $C_i$, representing the exact integer $S=\sum_i\rho^i C_i$, where $S\equiv y\pmod q$ for output residue $y$.
Carry propagation converts the accumulation planes into an exact radix-$\rho$ representation of $S$.
Modular recovery then computes the canonical residue $S\bmod q\in[0,q)$, which is repacked for the next matrix-transform stage.

\subsection{Shoup Modular Multiplication}
\label{sec:fixed_operand}

Shoup multiplication accelerates repeated modular multiplication by a fixed operand by replacing division with a precomputed quotient.
For a fixed operand $a\in[0,q)$, Shoup precomputes
$s=\lfloor aB/q\rfloor$, where $B=2^w>q$.
Given $x\in[0,q)$, it computes $t=\lfloor xs/B\rfloor$ and $r=xa-tq$~\cite{harvey_Faster_2014}.
The result satisfies $0\leq r<2q$, so at most one conditional subtraction is required to obtain $xa\bmod q$.

The conventional Shoup representation stores two machine words $(a,s)$ per fixed coefficient, increasing the storage and data-movement costs of evaluation-key material containing many such coefficients.
Barrett and Montgomery reduction use representations that differ from this coefficient--quotient pair~\cite{barrett1987rsa_dsp,montgomery1985modular}.

\section{Motivation}
\label{sec:motivation}
\subsection{NTT Arithmetic and Axis-Transform Boundaries}
\label{sec:motivation_ntt}

Key‑switching invokes NTT operations repeatedly, while multiplication, rotation, and bootstrapping all rely on key‑switching.
NTT efficiency therefore affects the performance of these higher-level operations.

\begin{figure}[t]
  \centering
  \includegraphics[width=\columnwidth]{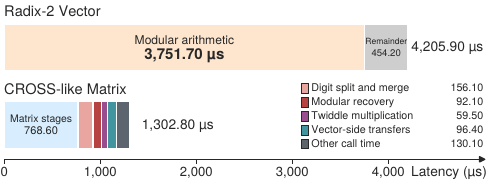}
  \caption{Vector modular arithmetic dominates NTT latency.
    \normalfont At \(N=65{,}536\) and 40 polynomial transforms. Remainder is the latency after replacing modular arithmetic with low-cost operations.}
  \Description{Two stacked bars share one latency axis. The Radix-2 Vector bar totals 4205.90 microseconds, of which modular arithmetic is 3751.70 and other Vector time 454.20. The CROSS-like Matrix bar totals 1302.80 microseconds: matrix stages 768.60, digit splitting and merging 156.10, modular recovery 92.10, twiddle modular multiplication 59.50, Vector-side transfers 96.40, and other call time 130.10. Each row total is printed next to its bar.}
  \label{fig:ntt_baseline}
\end{figure}

Figure~\ref{fig:ntt_baseline} shows that modular arithmetic dominates the execution time of the Vector-only Stockham radix-2 NTT.
In a controlled experiment with $N=65{,}536$ and 40 polynomials, preserving the loops, data movement, and reordering while replacing only the modular arithmetic with vector additions of zero reduces latency from $4205.90\,\mu s$ to $454.20\,\mu s$, a reduction of $89.20\%$.

For the same workload, our matrix-based NTT baseline using a CROSS-like mapping~\cite{tong_CROSS_2026_v4} achieves a latency of $1302.80\,\mu s$.
In the traced call of Figure~\ref{fig:ntt_baseline}, digit splitting and merging, modular recovery at axis endpoints, and inter-axis twiddle multiplication still account for $23.60\%$, while matrix stages and their internal transfers and waits account for $59.00\%$.
Further decomposition reduces matrix multiply--accumulate work but increases the overhead of modular recovery and representation conversion at intra-axis transform boundaries.
Matrix decomposition must therefore balance the savings in matrix computation against the additional work introduced between transforms.
\subsection{Limb-Scheduling Pressure}
\label{sec:motivation_limb}

\begin{figure}[b]
  \centering
  \includegraphics[width=\columnwidth]{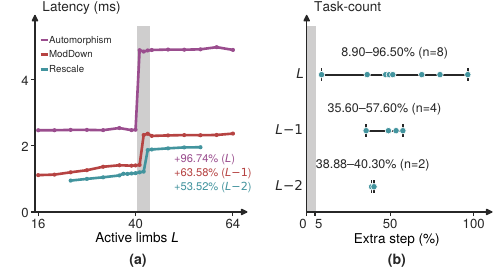}
  \caption{Whole-limb task assignment causes latency jumps at task-count boundaries.
  \normalfont \(N=65{,}536\); (a) latency curves for three kernel variants; (b) extra latency across task-count boundaries for 14 kernel variants.}
  \Description{The left panel plots latency in milliseconds against the active limb count for three kernel variants, which jump near 41, 42, and 43 limbs. The right panel groups 14 kernel variants by task-count rule, plotting each variant's extra step over the local trend as a dot strip with group ranges, counts, and the 5 percent criterion band.}
  \label{fig:limb_step}
\end{figure}

The active limb count $L$ varies across CKKS operations, while the worker count remains fixed.
Whole-limb task assignment produces an underfilled tail wave when the task count crosses worker capacity, causing a latency jump.
Figure~\ref{fig:limb_step}(a) shows jumps at $L=41$, $42$, and $43$ for kernel variants with $L$, $L-1$, and $L-2$ tasks, respectively.
Figure~\ref{fig:limb_step}(b) summarizes 14 surveyed NPUFHE kernel variants following these task-count rules, with jumps exceeding the $5\%$ criterion and ranging from $8.90\%$ to $96.50\%$.

Increasing task parallelism can also incur overhead: in our NTT control, replacing all whole-limb tasks with tile tasks increases latency by approximately $32.00\%$.
Mitigating tail-wave imbalance therefore requires controlling both the scope and granularity of splitting.

\subsection{Evaluation-Key Material Pressure}
\label{sec:motivation_keys}

In full-packing bootstrapping with $N=65{,}536$, $L=40$, $K=14$, and $\beta=3$, deduplicated rotation evaluation keys in paired Shoup representation still occupy $21.04\,\mathrm{GiB}$, with coefficients and precomputed quotients each accounting for half.
Parameters, transform constants, intermediate ciphertexts, and workspaces must also fit within the $32\,\mathrm{GiB}$ high-bandwidth memory (HBM), leaving insufficient capacity to keep all key material resident and requiring on-demand transfers from the host.
This motivates us to compress the coefficient representation to expand the set of resident material and reduce repeated transfers while preserving the online Shoup multiplication path.

\section{NPUFHE Design and Optimization}
\label{sec:npufhe_design_optimization}

\begin{figure}[b]
  \centering
  \includegraphics[width=0.98\columnwidth]{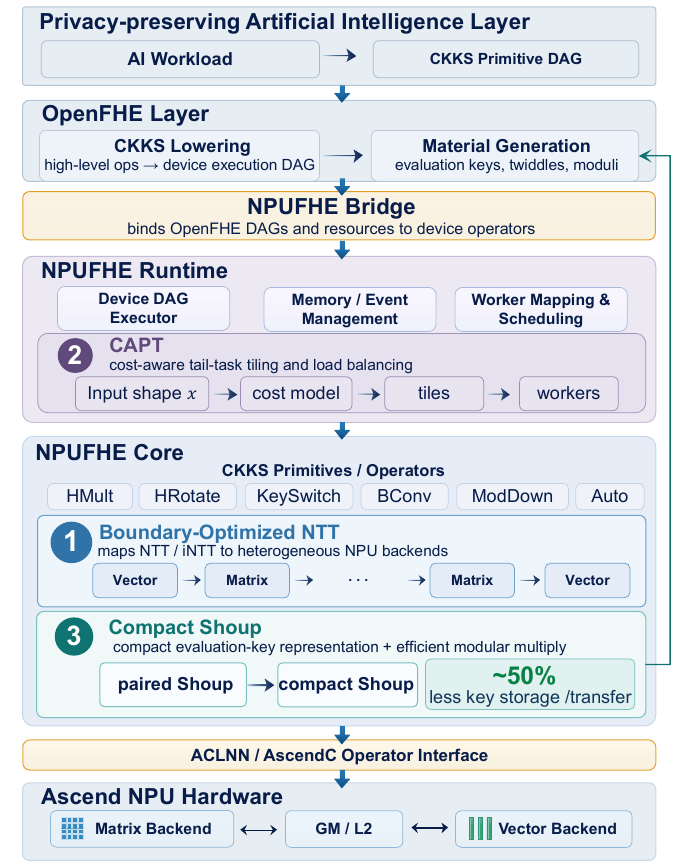}
  \caption{NPUFHE system overview.}
  \Description{System diagram of NPUFHE. A privacy-preserving AI layer lowers an example inference computation to a CKKS primitive DAG of PMult, HRotate, HAdd, HMult, Rescale, and PAdd. The OpenFHE layer performs CKKS lowering and material generation, and the NPUFHE bridge binds OpenFHE DAGs and resources to device operators. The NPUFHE runtime hosts the device DAG executor, memory and event management, and worker mapping and scheduling, including CAPT, which tiles tail tasks and balances load. The NPUFHE core implements the CKKS primitives, including the boundary-optimized NTT, which maps NTT and iNTT onto the Matrix and Vector backends, and the Compact Shoup representation of evaluation-key material. The ACLNN/AscendC operator interface connects the core to the Ascend NPU hardware and memory hierarchy.}
  \label{fig:system_stack}
\end{figure}

Figure~\ref{fig:system_stack} presents the architecture overview of \textsc{NPUFHE}, from encrypted applications to execution on Ascend devices.
OpenFHE represents computation as directed acyclic graphs (DAGs) and provides the required evaluation keys and constants.
The Bridge binds DAG tasks and their resources to device operators, while the Runtime schedules and executes these operators.
BO-NTT, CAPT, and Compact Shoup optimize Vector computation between matrix transform boundaries, tail load balance, and evaluation-key storage and transfer, respectively.

\subsection{Boundary-Optimized NTT}
\label{sec:bounded_ntt}
The two-axis form in Section~\ref{sec:ntt_in_ckks} expresses the NTT as a first-axis matrix transform, an inter-axis twiddle, and a second-axis matrix transform.
Let $N_1=d_1d_2$, with $d_1\ge d_2$.
We further split the first axis into two small matrix transforms, yielding an asymmetric decomposition:
\begin{equation}
 Y=\Bigl(\mathcal F_{d_2}\bigl(\mathcal F_{d_1}(X)\odot W_F\bigr)\odot W\Bigr)H\pmod q.
 \label{eq:ntt_first_axis_split}
\end{equation}
Here $\mathcal F_{d_1}$ and $\mathcal F_{d_2}$ act successively along the two factor dimensions of the first axis, and $W_F$ is the intra-axis twiddle between them; the second axis $H$ remains a direct matrix transform to avoid the additional intermediate layout reorder that further decomposition would introduce.

\begin{figure}[!b]
  \centering
  \includegraphics[width=\columnwidth]{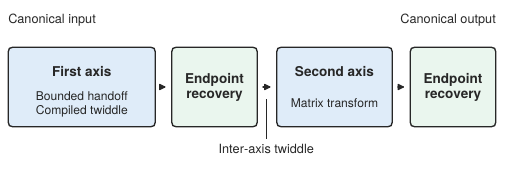}
  \caption{BO-NTT execution path.}
  \Description{Conceptual BO-NTT execution path. The first axis uses a bounded intermediate representation and compiled intra-axis twiddle. Endpoint recovery restores canonical residues after each axis, while the inter-axis twiddle remains explicit between the first endpoint and the second axis.}
  \label{fig:ntt_workload}
\end{figure}

We call the boundary and twiddle between the first axis's two small transforms \emph{intra-axis}, and the twiddle between the two axes \emph{inter-axis}.
Canonical residues are restored at the two \emph{axis endpoints}, after the first and second axes.
As Figure~\ref{fig:ntt_workload} shows, BO-NTT replaces intra-axis modular recovery with a six-digit radix-128 bounded intermediate representation, compiles $W_F$ and the digit weights into the second matrix's constants, and performs two-level folded modular recovery at axis endpoints.

\subsubsection{Bounded Intermediate Representation}
\label{sec:ntt_handoff}
\textsc{NPUFHE} uses 28-bit RNS moduli, so a canonical residue has the exact four-digit representation $x=\sum_{e=0}^{3}x_e\rho^e$ with $\rho=128$, where $0\leq x_e<\rho$.
Matrix constants are weighted by digit position and represented in the same radix, allowing the Matrix backend to use signed INT8 inputs and INT32 accumulation.

For an input $x_r=\sum_{e=0}^{p-1}x_{r,e}\rho^e$, we precompute the canonical weighted constants
$A^{(e)}_{j,r}=(\rho^e A_{j,r})\bmod q$.
The resulting weighted integer satisfies
\[
 T_j=\sum_{r,e}A^{(e)}_{j,r}x_{r,e}
 \equiv\sum_r A_{j,r}x_r\pmod q.
\]
Thus, the matrix computation preserves an exact integer congruent to the target matrix product modulo $q$; its integer value generally differs from the unreduced original dot product.

For reduction length $k$, radix-128 multiplication gives the bound
$T_j\le pk(\rho-1)^2\sum_{i=0}^{3}\rho^i$.
The first small transform is chosen so that its output fits exactly in six radix-128 digits:
$T_j=\sum_{e=0}^{5}D_{j,e}\rho^e$, with $0\le D_{j,e}<\rho$.
The Vector backend preserves this integer in the bounded intermediate representation, and the next matrix transform consumes all six digits.
This handoff eliminates intra-axis modular recovery and re-decomposition of canonical residues.

\subsubsection{Compiled Twiddle}
\label{sec:ntt_twiddle}
For each group of the second small transform, fix the remaining factor coordinates and let $A_2$ denote the transform matrix and $w_r$ the intra-axis twiddle at input position $r$.
We precompute $G_{j,r}=(A_2)_{j,r}w_r\bmod q$ and
$G^{(e)}_{j,r}=\rho^e G_{j,r}\bmod q$ for $0\le e<6$.
The integer $T_r=\sum_{e=0}^{5}D_{r,e}\rho^e$ supplied by the first small transform is congruent modulo $q$ to its canonical output $z_r$.
Thus, the output of the second small transform satisfies
$y_j\equiv\sum_{r,e}G^{(e)}_{j,r}D_{r,e}
\equiv\sum_r(A_2)_{j,r}w_rz_r\pmod q$.

The weighted constant matrices incorporate both the intra-axis twiddle and digit weights.
Their radix-128 representation retains INT8 inputs and INT32 accumulation while eliminating intra-axis modular recovery and standalone Vector-side twiddle modular multiplication.
The Vector backend still computes the inter-axis twiddle $W$ explicitly after the first axis completes, because folding it into the second-axis matrix would introduce constants dependent on the first-axis output index, increasing storage and reducing reuse.

\subsubsection{Two-Level Folded Modular Recovery}
\label{sec:ntt_endpoint}
Modular recovery remains only at the two axis endpoints, after the first and second axes.
For a matrix stage with $p$ input digits and reduction length $k$, each INT32 accumulation plane is bounded by $pk(\rho-1)^2$, and its weighted endpoint integer satisfies
$S\le pk(\rho-1)^2\sum_{i=0}^{3}\rho^i$.
The selected decomposition respects the INT32 accumulation range and the range required by endpoint recovery.

Two-level folding expresses the endpoint integer as
$S=s_{\mathrm{lo}}+2^{28}s_{\mathrm{hi}}$, giving
\[
 S\bmod q=(s_{\mathrm{lo}}+\kappa s_{\mathrm{hi}})\bmod q,
 \qquad \kappa=2^{28}\bmod q.
\]
This identity replaces digit-by-digit Horner recovery with component merging and one fixed-operand modular multiplication of the high part.
The endpoint ranges ensure $0\le s_{\mathrm{lo}}<2^{28}$ and $0\le s_{\mathrm{hi}}<q$.
Since $2^{27}<q<2^{28}$, the low part is below $2q$.
Combining the two parts modulo $q$ restores a canonical residue in $[0,q)$.
Together with the congruence preserved by each matrix stage, recovery only at the two axis endpoints preserves equivalence to the original NTT modulo $q$.

\subsection{Cost-Aware Parallel Tiling}
\label{sec:catp}

CAPT tiles only tail tasks and selects the granularity using a cost model calibrated offline, trading off load-balancing gains against fine-grained execution efficiency.

\subsubsection{Legal Output Tiles}

For a baseline task with output set \(G\), CAPT constructs a disjoint partition \(G=\biguplus_{k=0}^{g-1}G_k\), where \(g\) is the number of tiles, so that each tile independently produces complete outputs and satisfies three conditions.
(1) \textbf{Computational closure}: complete all reductions and postprocessing without splitting the reduction dimension of any output.
(2) \textbf{Exclusive outputs}: preserve global coordinates and data layout, allowing shared read-only inputs and constants without write conflicts or cross-tile reductions.
(3) \textbf{Physical executability}: satisfy alignment requirements for matrix computation, vector access, and data movement, with accumulation state and temporary buffers fitting in on-chip memory.

Accordingly, NTT/iNTT variants are tiled over output-column blocks of the subtransforms, with each block preserving the complete reduction. 

Let kernel variant \(e\) generate \(M_e(x)=hP+r\) baseline tasks for input shape \(x\), where \(P\) is the number of workers.
Here \(h=\lfloor M_e(x)/P\rfloor\) is the number of full execution waves, and \(r=M_e(x)\bmod P\) is the number of tail-wave tasks, with \(0\le r<P\).
CAPT retains \(hP\) whole tasks and their original mapping, and partitions each of the remaining \(r\) tasks into \(g\) legal output tiles to expose more tail-wave parallelism.

\subsubsection{Cost-Aware Granularity Selection}

Tail-task tiling reduces the load of the busiest worker, but its benefit depends on the tiling granularity.
Increasing the split factor \(g\) exposes more parallelism while increasing data-movement and execution-organization costs.

Taking BO-NTT at \(N=2^{16}\) and \(L{=}41\) as an example, whole-limb task assignment gives one worker two limb tasks.
CAPT retains the first 40 whole tasks and partitions the last task along output columns into \(g\in\{1,2,4,8,16\}\) legal tiles, reducing the busiest worker's equivalent workload to \(1+1/g\).
Each tile preserves the complete reduction and original data layout.

The cost model considers three categories.
\textbf{Load imbalance} captures the workload of the busiest worker and shared Matrix compute-resource group.
\textbf{Data movement} captures transfer volume and organization.
\textbf{Execution organization} captures the effects of task partitioning and kernel calls.
These costs jointly describe the tradeoff between reducing tail-wave imbalance and increasing fine-grained execution overhead.

Offline calibration uses measured kernel invocation latencies to estimate shared costs and account for differences among kernel variants.
At runtime, CAPT evaluates the legal candidate granularities and selects the one with the lowest predicted kernel invocation latency.

\subsubsection{Tail-Task Tiling}
\begin{figure}[t]
  \centering
  \includegraphics[width=\columnwidth]{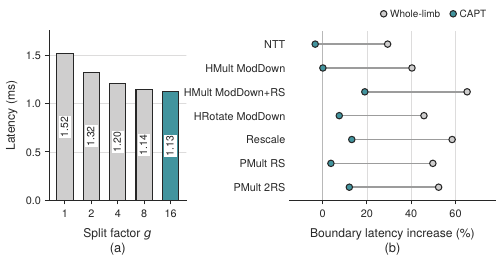}
  \caption{CAPT mitigates tail-wave latency on Ascend 910B4. 
  \normalfont (a) NTT kernel invocation latency at \(L=41\) under different split factors \(g\), where CAPT selects \(g=16\). (b) Boundary latency increase for seven kernel variants under whole-limb assignment and CAPT.}
  \Description{Two panels. Left: NTT kernel invocation latency falls from 1.52 milliseconds at split factor 1 to 1.13 milliseconds at split factor 16. Right: for seven kernel variants, the boundary latency increase drops from 29.34 to 65.25 percent under whole-limb task assignment to between minus 3.36 and 19.04 percent with CAPT.}
  \label{fig:capt_boundary}
\end{figure}

Figure~\ref{fig:capt_boundary}(a) shows the effect of tiling granularity on this NTT instance: latency falls from 1.52\,ms at \(g{=}1\) to the measured optimum of 1.13\,ms at \(g{=}16\), while the benefit of increasing \(g\) from 8 to 16 has already diminished.
We further apply CAPT to forward-NTT kernel variants within other operators.
As shown in Figure~\ref{fig:capt_boundary}(b), when these kernel variants cross a task-wave boundary, the legal tiling configurations selected by the cost model match the measured optimum at each endpoint, reducing the latency increase from 29.34--65.25\% under whole-limb task assignment to $-3.36$--19.04\%.

\subsection{Compact Shoup Encoding}
\label{sec:compact_shoup}

Compact Shoup stores only the precomputed quotient $s$ instead of the pair $(a,s)$, reducing the storage of evaluation-key material from 8 to 4 bytes per coefficient.
The Vector backend recovers $a$ in the UB and immediately uses it in Shoup multiplication, without writing it back to global memory.

\paragraph{Quotient encoding and coefficient recovery.}
We use $B=2^{28}$ with $2^{27}<q<B$.
For $0\leq a<q$, preprocessing computes $s=\lfloor aB/q\rfloor$.
Given a legal code $s$, the coefficient $a$ lies in $[sq/B,(s+1)q/B)$.
Its width is $q/B<1$, so legality uniquely determines $a$.
With $\delta=B-q$, we recover it as
\[
 a=\lceil sq/B\rceil=s-\lfloor s\delta/B\rfloor.
\]
The quotient-only encoding preserves the coefficient exactly.

\paragraph{Fused recovery and multiplication.}
The consumer fuses coefficient recovery with Shoup multiplication and accumulation.
Recovery remains local to the UB, so the recovered coefficient does not create an additional global-memory representation.
The fused path uses 32-bit arithmetic and reuses intermediate work between recovery and quotient estimation.

For canonical $x$, the Shoup estimate $t=\lfloor xs/B\rfloor$ is at most one below the true quotient, giving $r=xa-tq\in[0,2q)$.
This residual bound allows the consumer to defer modular reduction across a bounded number of product terms.

\paragraph{Deferred reduction.}
The consumer processes $1\le\beta\le3$ product terms, where $\beta$ denotes the number of key-switch decomposition groups.
For canonical inputs $0\le x_i,z<q$, it produces a canonical result
\[
 R\equiv z+\sum_{i=0}^{\beta-1}x_i a_i\pmod q,
 \qquad 0\le R<q.
\]
The sum of up to three uncorrected residuals is below $6q$ with a zero initial value, or below $7q<2^{31}$ when including a canonical ciphertext term, so signed 32-bit accumulation is safe.
Reduction restores the canonical range at the consumption boundary.
The same deferred-reduction strategy applies to both paired and compact representations.

\subsection{System Integration and Execution Path}
\label{sec:ntt_dataflow}
\begin{figure}[ht]
  \centering
  \includegraphics[width=\columnwidth]{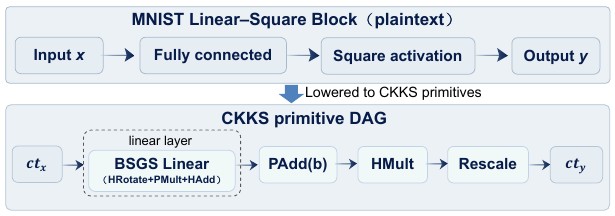}
  \caption{MNIST linear–square block lowered to a CKKS primitive DAG.
  \normalfont BSGS implements the linear layer, followed by bias addition, HMult, and Rescale.}
  \Description{Two stacked panels. Top: a plaintext MNIST linear--square block, in which input x passes through a fully connected layer and a square activation to output y. Bottom: the lowered CKKS primitive DAG, where the input ciphertext passes through a BSGS linear layer that expands into HRotate, PMult, and HAdd, then bias addition, homomorphic multiplication, and rescaling to the output ciphertext.}
  \label{fig:mnist_lowering}
\end{figure}

NPUFHE lowers high-level CKKS operations into an operator DAG that records inter-node data dependencies and node metadata, including the modulus basis, scale, representation domain, and encoding state.
Figure~\ref{fig:mnist_lowering} illustrates the lowering of a fully connected layer and square activation: baby-step giant-step (BSGS) implements the linear layer, followed by bias addition, HMult, and Rescale.
Where data dependencies and numerical bounds allow, NPUFHE further fuses adjacent physical kernels to reduce the materialization and movement of intermediate results in GM.
Fusion spans representation conversion, basis conversion, and arithmetic stages within composite CKKS operations.
It also extends across operator boundaries, reducing redundant transforms and intermediate data movement.

The Runtime coordinates execution of the fused DAG.
It uses node metadata to propagate changes in the modulus basis, scale, and RNS limb count, and allocates and reuses device memory according to data dependencies and buffer lifetimes.
Within the HBM budget, selected fixed operands remain resident on the device, while the remaining operands are prefetched from the host in chunks as needed.
Event dependencies coordinate data transfers, computation, and buffer reuse asynchronously, reducing host synchronization between nodes.
Task scheduling exposes parallel work across the Matrix, Vector, and data-movement pipelines.
After execution, the Bridge converts representation domains as needed and reconstructs the OpenFHE ciphertexts.

\section{Evaluation}
\label{sec:evaluation}

We evaluate \textsc{NPUFHE} at three levels: NTT, homomorphic operations, and encrypted applications. We compare it with existing solutions that follow a similar technical approach on various hardware platforms and parameter sets, and use same-platform ablations to analyze boundary-optimized NTT, CAPT, and Compact Shoup.

\subsection{Experimental Setup}
\label{sec:eval_setup}

We implement NTT and CKKS in NPUFHE, where the Vector unit operates on 28-bit residues, the Cube kernel uses 7-bit digit decomposition, rescaling adopts the double-prime scheme~\cite{cheon_rns_2019,kim_Reduced_2022}, and bootstrapping follows the Han--Ki scheme~\cite{han_Better_2020}.
All experiments are conducted on a single Ascend 910B4 card with 20 Cube cores, 40 Vector cores, approximately 540.70 INT8 TOPS and 4.22 INT32 TOPS, 192 KB of UB, and 0.80 TB/s HBM bandwidth. The host platform is a Kunpeng 920 with g++ 11.4.0, running CANN 8.3.RC1.alpha001 and OpenFHE 1.5.1.

\begin{table}[htbp]
  \centering\footnotesize
  \setlength{\tabcolsep}{3pt}
  \caption{FHE parameter sets for the NTT and homomorphic-operation comparison, all at 128-bit security.}
  \label{tab:eval_he_params}
  \begin{tabular*}{\columnwidth}{@{\extracolsep{\fill}}lcccccc@{}}
    \toprule
    & Set-A & Set-B & Set-C & Set-D & Set-E & Set-F \\
    \midrule
    $N$ & $2^{12}$ & $2^{13}$ & $2^{14}$ & $2^{15}$ & $2^{16}$ & $2^{16}$ \\
    $L$ & 2 & 6 & 14 & 24 & 34 & 36 \\
    $log(QP)$ & 108 & 217 & 437 & 704 & 974 & 1{,}337 \\
    \bottomrule
  \end{tabular*}
\end{table}

For the NTT and homomorphic-operation comparisons, we use the parameter sets listed in Table~\ref{tab:eval_he_params}, where $N$ denotes the polynomial degree, $L$ the number of RNS limbs, and $log(QP)$ the total bit length of the ciphertext modulus and special prime.

Unless stated otherwise, all measurements use 20 warm-up runs and 100 repetitions and report the median. NPU results are host--device end-to-end wall-clock times, and the CPU baseline is the process wall-clock time of OpenFHE on one physical core with one thread.

\textbf{Application configurations.}
Here $\beta$ is the number of key-switch decomposition groups, $K$ the number of auxiliary P-basis limbs, and $D=L+K$ the extended basis size.
Bootstrapping uses $N=65{,}536$, $L=46$, $\beta=3$, and $K=16$, with full packing and one bootstrap.
MNIST uses $N=16{,}384$, $L=12\rightarrow2$, $\beta=3$, and $K=4$, packing eight images per ciphertext without bootstrapping.
ResNet-20 inference on the NPU uses $N=65{,}536$, $L=46$, $K=16$, and $\beta=3$.
The three residual stages use sparse packing in $16{,}384/8{,}192/4{,}096$ logical slots, respectively, and each complete inference executes 19 bootstraps.

\subsection{Comparison with Existing Solutions}
\label{sec:eval_comparison}

We compare NPUFHE with representative solutions on mainstream platforms: FAB~\cite{agrawal_FAB_2023} on FPGA, Over 100x~\cite{jung_100x_2021}, TensorFHE~\cite{fan_TensorFHE_2023}, and WarpDrive~\cite{fan_WarpDrive_2025} on GPU, and CROSS~\cite{tong_CROSS_2026_v4} on TPU, as the general-purpose baseline, we measure OpenFHE~\cite{openfhe} on one physical Kunpeng-920 core with one OpenMP thread.
Table~\ref{tab:eval_reported} summarizes the hardware platforms and clock frequencies of the compared solutions.

\begin{table}[H]
  \centering
  \footnotesize
  \caption{Hardware platforms of the compared solutions.}
  \label{tab:eval_reported}
  \begin{tabular*}{\columnwidth}{@{\extracolsep{\fill}}lllr@{}}
    \toprule
    Scheme & Type & Platform & Clock (GHz) \\
    \midrule
    OpenFHE & CPU & Kunpeng-920 (1 core) & 2.60 \\
    FAB & FPGA & Alveo U280 & 0.30 \\
    Over 100x & GPU & Tesla V100 16GB & 1.20 \\
    TensorFHE & GPU & A100-SXM-40GB & 1.41 \\
    WarpDrive & GPU & A100-PCIE-80GB & 1.41 \\
    CROSS & TPU & TPU v4-8 (4 chips) & 1.05 \\
    \bottomrule
  \end{tabular*}
\end{table}

Except for OpenFHE, all performance data are taken from the original papers to keep the cross-platform comparison traceable. Tables~\ref{tab:ntt_set_throughput} and~\ref{tab:eval_he} display throughput and latency rounded to integers.

Table~\ref{tab:ntt_set_throughput} reports NTT throughput (KNTT/s) on Sets A--E. On one Ascend 910B4, NPUFHE reaches 1023.44/609.32/334.44/127.62/98.30 KNTT/s: 77.27--184.22$\times$ that of OpenFHE (one forward \texttt{SwitchFormat()} per sample on one pinned Kunpeng-920 core), 1.12--2.04$\times$ that of TensorFHE, up to 4.46$\times$ that of CROSS v4-8 (four chips), and about 2.00$\times$ that of FAB.
As $N$ increases from $2^{12}$ to $2^{14}$, NPUFHE retains 32.70\% of its throughput, whereas CROSS v4-8 retains 5.80\%, so NPUFHE's relative throughput advantage grows with scale.

\subsubsection{NTT Performance Comparison}
\label{sec:eval_ntt}

\begin{table}[htbp]
  \centering
  \caption{NTT throughput (KNTT/s) across Sets A--E. }
  \label{tab:ntt_set_throughput}
  \scriptsize
  \setlength{\tabcolsep}{2.5pt}
  \begin{tabular*}{\columnwidth}{@{\extracolsep{\fill}}lrrrrr@{}}
    \toprule
    Scheme & Set-A & Set-B & Set-C & Set-D & Set-E \\
    \midrule
    \multicolumn{6}{@{}l}{\textit{Reported values (original papers)}} \\
    FAB & -- & -- & 167 & -- & -- \\
    TensorFHE & 910 & 450 & 209 & 99 & 48 \\
    CROSS v4-8 & 1{,}284 & 323 & 75 & -- & -- \\
    \midrule
    \multicolumn{6}{@{}l}{\textit{Measured in this work}} \\
    OpenFHE (CPU) & 13 & 6 & 3 & 1 & 1 \\
    Vector (control) & 192 & 90 & 43 & 20 & 10 \\
    CROSS-like (control) & 587 & 490 & 281 & 128 & 46 \\
    \textbf{NPUFHE} & \textbf{1{,}023} & \textbf{609} & \textbf{334} & \textbf{128} & \textbf{98} \\
    \midrule
    \multicolumn{6}{@{}l}{\textit{Speedup of NPUFHE}} \\
    vs.\ OpenFHE (CPU) & 77.27$\times$ & 101.50$\times$ & 125.19$\times$ & 108.59$\times$ & 184.22$\times$ \\
    vs.\ TensorFHE & 1.12$\times$ & 1.35$\times$ & 1.60$\times$ & 1.29$\times$ & 2.04$\times$ \\
    \bottomrule
  \end{tabular*}
  \begin{minipage}{\columnwidth}
    \normalfont\scriptsize
    ``--'' is not reported. OpenFHE is measured on one physical Kunpeng-920 core with one OpenMP thread. Vector and CROSS-like are our own control implementations on Ascend 910B4.
  \end{minipage}
\end{table}

\begin{figure}[htbp]
  \centering
  \includegraphics[width=\columnwidth]{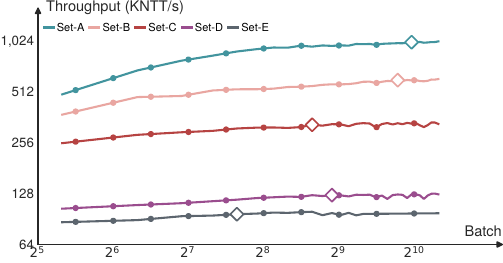}
  \caption{NTT batch scaling across Sets A--E on one Ascend 910B4}
  \Description{A single panel. Five curves give throughput in KNTT/s against the batch counted in active RNS limbs per call, for Sets A to E. All curves saturate with batch; the higher sets saturate at fewer limbs and at lower throughput. Filled circles sit under the power-of-two ticks and open diamonds mark 97 percent of each measured peak.}
  \label{fig:ntt_throughput}
\end{figure}

Compared with the pure-Vector and CROSS-like controls on the same NPU, our NTT also achieves clear gains: it outperforms the pure-Vector control at all five polynomial degrees and CROSS-like at four.
Although NPUFHE's standalone NTT throughput on Set-E is 98.30 KNTT/s, lower than the 468 KNTT/s reported by WarpDrive, this approximately $4.76\times$ gap at the individual-operator level does not carry over proportionally to complete homomorphic operations. For HMult, NPUFHE's latency is only 4.69\% higher than WarpDrive's.
See Section~\ref{sec:eval_kernels} for the corresponding analysis.

As shown in Figure~\ref{fig:ntt_throughput}, NTT throughput rises with batch and approaches saturation. Larger parameter sets saturate at smaller batches but achieve lower peak throughput.

\subsubsection{Homomorphic-Operation Comparison}
\label{sec:eval_kernels}
On Set-E, NPUFHE's HMult latency is 4.69\% higher than WarpDrive's, while achieving speedups of 1.24$\times$ and 2.45$\times$ over Over 100x and the TensorFHE reproduction, respectively. On Set-F, its HMult latency is 4.53\% higher than CROSS's (Table~\ref{tab:eval_he}).

NPUFHE's standalone Rescale remains slower than prior implementations (Table~\ref{tab:eval_he}).
However, Rescale typically follows multiplication in CKKS, so we jointly optimize HMult--Rescale, fusing adjacent stages to eliminate redundant transforms and intermediate materialization.
On Set-E, the fused chain takes 4{,}583\,$\mu$s, only slightly above the 4{,}485\,$\mu$s for standalone HMult.

\begin{table}[htbp]
  \centering
  \footnotesize
  \setlength{\tabcolsep}{2.5pt}
  \caption{Homomorphic-operation latency ($\mu$s). Parameter sets are defined in Table~\ref{tab:eval_he_params}.}
  \label{tab:eval_he}
  \begin{tabular*}{\columnwidth}{@{\extracolsep{\fill}}llrrrrr@{}}
    \toprule
    Scheme & Param. & HMult & HRotate & Rescale & HMult+RS & HAdd \\
    \midrule
    \multirow{3}{*}{TensorFHE-repl} & Set-C & 847 & 838 & 149 & 996$^{*}$ & 5 \\
    & Set-D & 2{,}893 & 2{,}876 & 355 & 3{,}248$^{*}$ & 11 \\
    & Set-E & 10{,}986 & 11{,}030 & 759 & 11{,}745$^{*}$ & 61 \\
    \multirow{3}{*}{Over 100x} & Set-C & 504 & 512 & 87 & 591$^{*}$ & 12 \\
    & Set-D & 1{,}642 & 1{,}667 & 181 & 1{,}823$^{*}$ & 21 \\
    & Set-E & 5{,}571 & 5{,}659 & 396 & 5{,}967$^{*}$ & 82 \\
    \multirow{3}{*}{WarpDrive} & Set-C & {277} & {273} & {45} & 322$^{*}$ & 5 \\
    & Set-D & {1{,}089} & {1{,}095} & {100} & 1{,}189$^{*}$ & 11 \\
    & Set-E & {4{,}284} & {4{,}341} & {241} & 4{,}525$^{*}$ & 61 \\
    CROSS & Set-F & 4{,}373 & 2{,}875 & 724 & 5{,}097 $^{*}$ & 43 \\
    \multirow{4}{*}{\textbf{NPUFHE}} & Set-C & \textbf{1{,}101} & \textbf{1{,}360} & \textbf{430} & \textbf{1{,}055} & \textbf{33} \\
    & Set-D & \textbf{2{,}416} & \textbf{2{,}820} & \textbf{760} & \textbf{2{,}422} & \textbf{32} \\
    & Set-E & \textbf{4{,}485} & \textbf{5{,}222} & \textbf{1{,}473} & \textbf{4{,}583} & \textbf{83} \\
    & Set-F & \textbf{4{,}571.3} & \textbf{5{,}140.0} & \textbf{1{,}258.0} & \textbf{4{,}655.5} & \textbf{84} \\
    \bottomrule
  \end{tabular*}
  \begin{minipage}{\columnwidth}
    \normalfont\scriptsize
    TensorFHE-repl denotes the TensorFHE reproduction reported by WarpDrive~\cite{fan_WarpDrive_2025}, not measurements from the original TensorFHE paper. Over 100x values are from its original paper~\cite{jung_100x_2021}. $^{*}$Sum of standalone HMult and Rescale latencies. NPUFHE HMult+RS is the measured fused-chain latency.
  \end{minipage}
\end{table}

\subsubsection{Workload Comparison}
\label{sec:eval_workloads}
We evaluate CKKS bootstrapping, MNIST, and CIFAR-10 ResNet-20.

\textbf{CKKS Bootstrapping.}
\label{sec:eval_bootstrap}
We evaluate full-packing CKKS bootstrapping with matched parameters on the CPU and NPU. Both output ciphertexts on both platforms retain 22 Q-basis RNS limbs.
On one physical CPU core with one thread, OpenFHE bootstrapping has a median latency of 294.83\,s and decrypted precision of 10.60 bits.

A complete NPU bootstrap using real evaluation keys takes 0.95\,s and achieves 10.24 bits of decrypted precision, yielding a speedup of approximately 310$\times$ over the single-core CPU.
Section~\ref{sec:shoup_evaluation} further reports Compact Shoup's reduction in evaluation-key material storage.

\textbf{MNIST.}
\label{sec:eval_ml}
MNIST uses an Orion-style $784\!\rightarrow\!128\!\rightarrow\!128\!\rightarrow\!10$ network with three linear layers and two exact squares.
Only the input is encrypted, and each ciphertext packs eight images in feature-major order and all three linear layers use BSGS diagonal encoding.
We validate this homomorphic execution path end to end on 1{,}000 test images.
Both CPU and NPU decrypted predictions achieve \textbf{98.20\%} accuracy, matching the plaintext reference, and NPU inference achieves a speedup of approximately \textbf{28$\times$} over the single-core CPU.

\begin{table}[htbp]
  \centering
  \scriptsize
  \setlength{\tabcolsep}{2pt}
  \caption{Complete CPU/NPU execution comparison for the end-to-end workloads.}
  \label{tab:e2e}
  \begin{tabular*}{\columnwidth}{@{\extracolsep{\fill}}llrrr@{}}
    \toprule
    Workload & System & Latency (s) & Precision (bits) & Acc. / $\Delta$Acc. (pp) \\
    \midrule
    \multirow{2}{*}{CKKS Bootstrap}
      & CPU & 294.83 & 10.60 & -- \\
      & \textbf{NPUFHE} & \textbf{0.95} & \textbf{10.24} & -- \\
    \midrule
    \multirow{2}{*}{MNIST}
      & CPU & 0.56 & 26.96 & 98.20\% / 0.00 \\
      & \textbf{NPUFHE} & \textbf{0.02} & \textbf{26.58} & \textbf{98.20\% / 0.00} \\
    \midrule
    \multirow{2}{*}{ResNet-20}
      & CPU & 810.70 & 6.91 & 92.21\% / $-0.20$ \\
      & \textbf{NPUFHE} & 9.87 & 6.43 & \textbf{92.21\%} / $\mathbf{-0.20}$ \\
    \bottomrule
  \end{tabular*}
  \begin{minipage}{\columnwidth}
    \normalfont\scriptsize
    Precision (bits) is $-\log_2\epsilon_{\max}$, where $\epsilon_{\max}$ is the maximum absolute error between decrypted and reference outputs. $\Delta$Acc. is the accuracy change relative to the plaintext reference on the same test set, in percentage points. Plaintext-reference accuracy is 98.20\% for MNIST and 92.41\% for ResNet-20.
  \end{minipage}
\end{table}

\textbf{ResNet-20.}
CIFAR-10 ResNet-20 evaluates deep encrypted inference with frequent bootstrapping.
We construct the NPU encrypted inference pipeline based on the fixed network architecture of LowMemoryFHEResNet20, executing 19 bootstraps in each complete inference.
The activation is a degree-59 Chebyshev polynomial on $[-1,1]$, whose evaluation, including the final scale normalization, consumes 7 logical levels.
The network consists of a stem and three residual stages, each containing three residual blocks, with channels of 16, 32, and 64 and spatial sizes of 32, 16, and 8, respectively. Global average pooling produces 10 output logits.
Only the input data are encrypted, while the convolutional, batch-normalization, and fully connected layer parameters remain in plaintext. The client decrypts the final output logits.

On the CPU, end-to-end encrypted inference of a single CIFAR-10 image takes \textbf{810.70\,s}, whereas NPUFHE reduces this latency to \textbf{9.87\,s}.
Both CPU and NPU encrypted inference achieve \textbf{92.21\%} classification accuracy, compared with \textbf{92.41\%} for the plaintext reference, a decrease of only \textbf{0.20 percentage points}.

\
\subsection{Ablation Analysis}
\label{sec:eval_ablation}
We use same-platform ablations to evaluate Boundary-Optimized NTT, Compact Shoup encoding, and CAPT.

\subsubsection{Boundary-Optimized NTT Ablation}
\label{sec:eval_ntt_ablation}
\begin{figure}[t]
  \centering
  \includegraphics[width=\columnwidth]{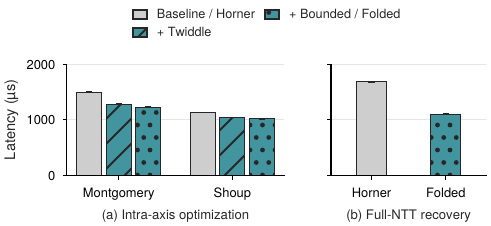}
  \caption{Complete-NTT latency ablations. \normalfont (a) Intra-axis optimization. (b) Horner versus folded endpoint recovery in a separate matched experiment. Whiskers span the minimum and maximum round medians.}
  \Description{Two panels compare complete NTT latency in microseconds on zero-based axes. Panel a shows baseline, twiddle compilation, and added bounded handoff for Montgomery and Shoup. Panel b compares complete NTT execution with Horner or folded recovery at both endpoints. Whiskers show the range of round medians.}
  \label{fig:ntt_four_factor}
\end{figure}

\textbf{Ablation setup.}
At $N=2^{16}$ and $L=40$, we separately evaluate twiddle compilation, bounded intermediate representation, and endpoint modular recovery (Figure~\ref{fig:ntt_four_factor}).
Each comparison holds the matrix decomposition, task mapping, 28-bit moduli, canonical input/output representations, and frequency order fixed. CAPT is disabled in all experiments.

\textbf{Intra-axis boundary optimization.}
Compiling the intra-axis twiddle into the matrix reduces complete NTT latency by 14.16\% under Montgomery and 7.65\% under Shoup (Figure~\ref{fig:ntt_four_factor}(a)).
On this basis, replacing intra-axis modular recovery and four-digit re-de\-com\-po\-si\-tion with a six-digit bounded intermediate representation further reduces latency by 4.21\% and 2.54\%, respectively.

\textbf{Axis-end modular recovery.}
In a separate matched Shoup experiment, we use the same full-width Shoup multiplication primitive to compare complete NTT execution with Horner recovery or folded recovery at both endpoints.
Folding both endpoints reduces complete NTT latency by 35.12\% (Figure~\ref{fig:ntt_four_factor}(b)).

\subsubsection{Compact Shoup Encoding Ablation}
\label{sec:shoup_evaluation}

\begin{figure}[b]
  \centering
  \includegraphics[width=\columnwidth]{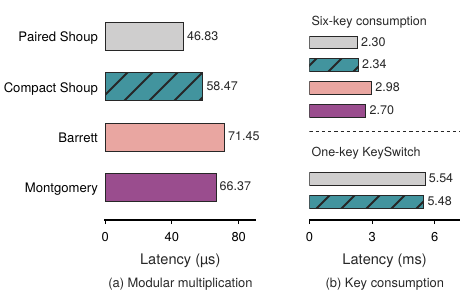}
  \caption{Compact Shoup comparisons at $L=40$. \normalfont (a) Four-arm canonical modular multiplication. (b) Matched six-key consumption; below the separator is a separate one-key complete resident KeySwitch. Colors identify the same methods across panels.}
  \Description{Two horizontal bar-chart panels. Panel a compares paired Shoup, Compact Shoup, Barrett, and Montgomery at 46.83, 58.47, 71.45, and 66.37 microseconds. Panel b compares the matched six-key consumer at 2.30, 2.34, 2.98, and 2.70 milliseconds, respectively. A dashed line separates the separate one-key complete resident KeySwitch pair, at 5.54 milliseconds for paired Shoup and 5.48 milliseconds for Compact Shoup.}
  \label{fig:compact_shoup_ablation}
\end{figure}

\textbf{Online modular multiplication.}
We compare the modular-multiplication latency of paired Shoup, Compact Shoup, Barrett, and Montgomery in Figure~\ref{fig:compact_shoup_ablation}(a).
Compact Shoup incurs 24.88\% higher latency than paired Shoup, but achieves 1.22$\times$ and 1.14$\times$ speedups over Barrett and Montgomery, respectively.

\textbf{Key-consumption ablation.}
Key consumption amortizes the single-product latency penalty by fusing recovery, multiplication, and accumulation.
We generate six real rotation keys for steps 1--6 in the same CKKS context to evaluate coefficient recovery and key consumption with Compact Shoup (Figure~\ref{fig:compact_shoup_ablation}(b)). The configuration uses $N=65{,}536$, $L=40$, $K=14$, and $\beta=3$. The four consumer arms use the same synthetic canonical decomposition digits, canonical outputs, and tile candidates. Montgomery keys are encoded offline. Timing covers the key inner product, accumulation, and final reduction.
In the matched four-arm consumer comparison, Compact takes 2.34\,ms versus 2.30\,ms for paired, a net penalty of 2.13\%, while achieving 1.27$\times$ and 1.15$\times$ speedups over Barrett and Montgomery.
Compact, Barrett, and Montgomery each require 486.00\,MiB of key representation, versus 972.00\,MiB for paired.
A separate single-key complete resident KeySwitch takes 5.48\,ms with Compact and 5.54\,ms with paired.

\textbf{Full-bootstrap system ablation.}
For the same full-packing $L=40$ bootstrapping configuration, Compact Shoup reduces the representation size of the rotation evaluation-key material after stage-specific projection and deduplication from 21.04\,GiB to 10.52\,GiB.
The smaller representation allows all four root rotation-key groups to remain resident, up from two, and reduces host-to-device material uploads per bootstrap from 16.85\,GiB to 2.72\,GiB, an 83.85\% reduction.
Under the same HBM budget, the P50 latency of a complete bootstrap after preparation decreases from 1.293\,s to 0.788\,s, yielding a $1.64\times$ speedup.

\subsubsection{CAPT Ablation}
\label{sec:catp_evaluation}

\begin{figure}[t]
  \centering
  \includegraphics[width=\columnwidth]{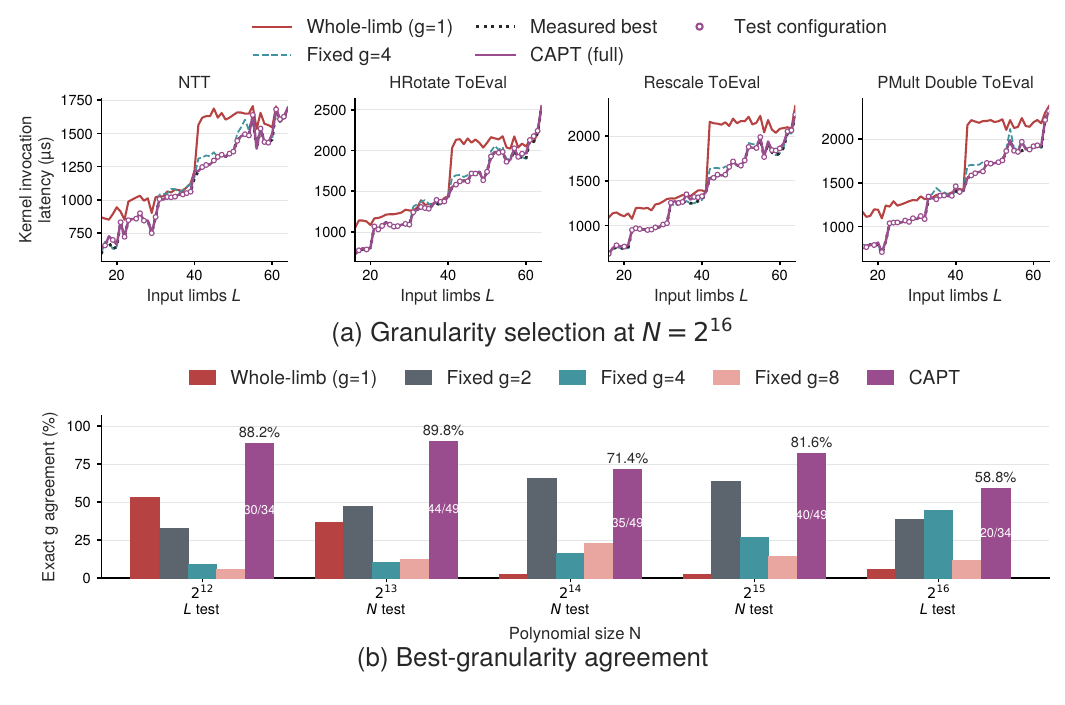}
  \caption{CAPT granularity selection. \normalfont (a) Median kernel invocation latency for four representative kernel variants at \(N=2^{16}\), comparing whole-limb assignment (\(g{=}1\)), fixed \(g{=}4\), the cost model's selection, and the fastest measured candidate. Hollow circles mark test configurations excluded from calibration. (b) Cross-\(N\) best-granularity agreement of the cost model and fixed \(g\in\{1,2,4,8\}\) policies on test configurations excluded from calibration.}
  \Description{Two vertically stacked panels. Panel a contains four latency plots for NTT, HRotate ToEval, Rescale ToEval, and PMult Double ToEval as input limb count increases from 16 to 64. Panel b compares exact agreement with the measured optimal granularity for four fixed policies and the cost model at five polynomial sizes. The cost model matches 30 of 34 test configurations at unseen L values at N4096, 44 of 49 test configurations at unseen N values at N8192, 35 of 49 at N16384, 40 of 49 at N32768, and 20 of 34 test configurations at unseen L values at N65536.}
  \label{fig:capt_ablation}
\end{figure}

We evaluate CAPT's granularity selection on 14 kernel variants exhibiting stepwise latency at \(N=2^{12}\)--\(2^{16}\) and \(L=16\)--64.
The sweep covers 3,430 kernel--configuration pairs, with candidate split factors \(g\in\{1,2,4,8,16\}\).

Figure~\ref{fig:capt_ablation}(a) shows the results for four representative kernel variants at \(N=2^{16}\).
Whole-limb assignment (\(g=1\)) produces pronounced stepwise latency near task-wave boundaries.
A fixed \(g=4\) alleviates load imbalance for some configurations but cannot accommodate all inputs.
CAPT dynamically selects \(g\) according to the kernel variant and input configuration, keeping the selected configuration's kernel invocation latency close to that of the fastest measured candidate across much of the sweep.
Compared with \(g=1\), CAPT reduces the mean kernel invocation latency of NTT, HRotate ToEval, Rescale ToEval, and PMult Double ToEval by 11.27\%, 10.86\%, 15.41\%, and 15.36\%, respectively.

We further ablate the cost model underlying the selections in Figure~\ref{fig:capt_ablation}(a) to examine the contribution of each cost component to CAPT's granularity selection.
We compare three configurations of the cost model, using the same calibrated \(g=1\) baseline latency and different components for relative scheduling costs.
The load-only configuration considers only the loads of the busiest worker and paired resource groups.
The shared configuration additionally accounts for call counts, data-movement pressure, and whole-task counts. The full configuration adds kernel-specific corrections.
All parameters are fitted only on the calibration set.

The three configurations have mean selection losses of 12.338\%, 0.499\%, and 0.456\%, respectively.
Selection loss is defined as
\begin{equation}
  100\left(T_{\mathrm{selected}}/T_{\mathrm{best}}-1\right),
  \label{eq:capt_selection_loss}
\end{equation}
measuring the performance loss of the selected configuration relative to the fastest measured configuration in the same candidate set.
These results show that modeling load alone is insufficient for accurate granularity selection.
Adding call and data-movement costs substantially reduces mean selection loss, while kernel-specific corrections provide a further small reduction.

Figure~\ref{fig:capt_ablation}(b) evaluates cross-\(N\) generalization of the cost model, independently calibrated for forward NTT with one parameter set shared across \(N\).
We calibrate the cost model on 15 \(L\) values at each of \(N=2^{12}\) and \(N=2^{16}\), freeze the parameters, and evaluate \(N=2^{12}\)--\(2^{16}\) across all 49 integer limb counts \(L=16\)--64 with candidate granularities \(g\in\{1,2,4,8\}\).
Each calibration size has 34 \(L\) values excluded from calibration, and all configurations at \(N=2^{13},2^{14},2^{15}\) are excluded from calibration.

Across the 147 cross-\(N\) test configurations at the three intermediate sizes, the cost model achieves 80.95\% exact agreement with the measured-optimal granularity and 1.14\% mean selection loss, compared with 58.50\% and 2.26\% for fixed \(g=2\), respectively.
Within the measured forward-NTT configurations, one shared parameter set therefore improves granularity selection across polynomial sizes with low mean selection loss, without per-\(N\) fitting.

\section{Related Work}
\label{sec:related_work}

\textbf{GPU.}
Over 100x reduces global memory traffic during bootstrapping through kernel fusion~\cite{jung_100x_2021}.
TensorFHE maps the NTT onto Tensor Cores and performs the main multiply--accumulate work with low-precision decomposition~\cite{fan_TensorFHE_2023}. WarpDrive further optimizes transform decomposition and on-chip residency~\cite{fan_WarpDrive_2025}, and Neo recasts scalar and element-wise multiplications as matrix multiplications and tailors the data layout to matrix computation~\cite{jiao_Neo_2025}.
TensorFHE+ extends this line of work by decomposing CKKS kernels into vector and matrix operations, vectorizing modular arithmetic, and optimizing data layouts~\cite{sun_TensorFHEPlus_2026}.
NPUFHE further reduces the auxiliary arithmetic between matrix computations: explicit integer bounds let intermediate results enter the next matrix stage without first being recovered to a canonical modular representation, and the intra-axis twiddle is absorbed into the next constant matrix.
Matrix acceleration therefore not only raises multiply--accumulate throughput but also removes one intra-axis modular recovery and the standalone intra-axis twiddle multiplication.

\textbf{FPGA.}
HEAX builds pipelined datapaths for the NTT and homomorphic computation~\cite{riazi_HEAX_2020}. FAB and Poseidon extend acceleration to bootstrapping~\cite{agrawal_FAB_2023,yang_Poseidon_2023}, and FAST designs a memory-efficient datapath for the homomorphic linear transformation that significantly reduces off-chip ciphertext access~\cite{xu_FAST_FPGA_2025}.
Flex-NTT unifies coefficient and twiddle-factor memory access patterns and reconfigures one hardware instance for NTT, inverse NTT, and point-wise multiplication~\cite{cheng_FlexNTT_2026}.
These designs gain efficiency from custom pipelines and buffering.
NPUFHE instead faces a task-granularity mismatch on fixed hardware: whole-limb task assignment favors locality, but stepwise latency increases appear once the task count crosses the worker-count boundary.
CAPT keeps whole-limb execution for the main tasks and splits only tail limbs with cost-aware selection, combining whole-limb locality with the load balance of fine-grained tasks.

\textbf{ASIC.}
F1 and CraterLake accelerate FHE through dedicated compute and storage organization~\cite{feldmann_F1_2021,samardzic_CraterLake_2022}. BTS makes bootstrapping a first-class citizen in accelerator design~\cite{kim_BTS_2022}, and ARK reduces the working set through runtime data generation and inter-operation key reuse~\cite{kim_ARK_2022}.
CROPHE explores spatial and temporal data pipelining and sharing across operators on a unified architecture, targeting on-chip reuse and reduced off-chip access~\cite{chen_CROPHE_2026}.
CROSS instead reuses a commercial TPU and uses Basis-Aligned Transformation (BAT) and Memory-Aligned Transformation (MAT) to optimize the low-precision arithmetic mapping and data layout~\cite{tong_CROSS_2026}.
NPUFHE also reuses commercial AI hardware, and it pushes the storage optimization down to the fixed-multiplier representation of evaluation-key material.
Compact Shoup keeps a single quotient encoding and recovers the coefficient when it is consumed, replacing the paired coefficient--quotient representation of Harvey~\cite{harvey_Faster_2014}. Fast Shoup modular multiplication therefore no longer needs two resident coefficient arrays, which reduces the storage and movement of evaluation-key material at the representation level.
On this basis, NPUFHE integrates these mechanisms into a complete CKKS execution through OpenFHE and validates the system with real evaluation keys and decrypted results.

\section{Conclusion}
\label{sec:conclusion}

In this paper, we present \textsc{NPUFHE}, an end-to-end CKKS execution system that turns Ascend NPUs already serving AI workloads into practical engines for encrypted computation.
It eliminates intra-axis modular recovery through a bounded intermediate representation and compiles intra-axis twiddle factors into matrix constants. 
CAPT mitigates tail load imbalance, and Compact Shoup halves the storage of evaluation-key material.
Integrated with OpenFHE, NPUFHE supports complete CKKS workloads ranging from homomorphic operations to bootstrapping and encrypted inference.
On one Ascend 910B4, it delivers up to $184.22\times$ the peak NTT throughput of OpenFHE on CPU, and further completes full-packing CKKS bootstrapping in 0.95\,s and encrypted CIFAR-10 ResNet-20 inference with bootstrapping in 9.87\,s per image.
These results establish deployed AI infrastructure as a practical substrate for privacy-preserving computing.
We will extend \textsc{NPUFHE} to multi-NPU execution and public-cloud deployment for large-scale encrypted AI services.

\bibliographystyle{ACM-Reference-Format}
\bibliography{references}
\end{document}